\documentclass[twocolumn,showpacs,showkeys,superscriptaddress,preprintnumbers,amsmath,amssymb,prl,longbibliography]{revtex4-1}
\usepackage{graphicx}
\usepackage{dcolumn}
\usepackage{bm}
\usepackage{amsmath}
\usepackage{color}

\begin{document}

\title{Steady Flow Dynamics during Granular Impact}

\author{Abram H. Clark}
\affiliation{Department of Physics \& Center for Nonlinear and Complex Systems, Duke University, Durham, North Carolina 27708, USA}
\affiliation{Department of Mechanical Engineering and Materials Science, Yale University, New Haven, Connecticut 06520, USA}
\author{Lou Kondic}
\affiliation{Department of Mathematical Sciences, New Jersey Institute of Technology, Newark, New Jersey 07102, USA}
\author{Robert P. Behringer}
\affiliation{Department of Physics \& Center for Nonlinear and Complex Systems, Duke University, Durham, North Carolina 27708, USA}

\begin{abstract}
We study experimentally and computationally the dynamics of granular flow during impacts, where intruders strike a collection of disks from above. In the regime where granular force dynamics are much more rapid than the intruder motion, we find that the particle flow near the intruder is proportional to the instantaneous intruder speed; it is essentially constant when normalized by that speed. The granular flow is nearly divergence-free and remains in balance with the intruder, despite the latter's rapid deceleration. Simulations indicate that this observation is insensitive to grain properties, which can be explained by the separation of time scales between intergrain force dynamics and intruder dynamics. Assuming there is a comparable separation of time scales, we expect that our results are applicable to a broad class of dynamic or transient granular flows. Our results suggest that descriptions of static-in-time granular flows might be extended or modified to describe these dynamic flows. Additionally, we find that accurate grain-grain interactions are not necessary to correctly capture the granular flow in this regime.
\end{abstract}
\date{\today}

\keywords{Granular materials, Granular flow, Impact}
\pacs{47.57.Gc, 81.05.Rm, 78.20.hb} 

\maketitle

What is the nature of force transmission and particle flow during dynamic intrusion into granular material? This question is fundamental
to a general understanding of dense granular flow, and a complete description would have many applications, such as biological or robotic locomotion over sand~\cite{Li2013,Aguilar2015} or impact into the surface of extraterrestrial bodies~\cite{Schwartz2014}. Moreover, the flow of grains during intrusion is part of a broad class of dense granular flows that are both rapid (\textit{i.e.}, large inertial number~\cite{daCruz2005}) and highly transient in time. The highly transient driving seemingly prohibits use of existing descriptions of dense granular flows~\cite{daCruz2005,Jop2006,Kamrin2012}, which are formulated for well developed cases (\textit{i.e.}, static-in-time after transients have settled, or quasi-static). Additionally, the large speeds and accelerations involved in this process raise important questions on how these flows should be considered computationally, either with a discrete element method (DEM)~\cite{cundall79,PicaCiamarra2004,pre12_impact} or from a continuum perspective~\cite{Dunatunga2015}.

In this Rapid Communication, we present experimental and computational
results on the flow of a 2D granular material around circular
intruders that are incident on a free granular bed at
speeds $v_0\leq 6 $~m/s. The main result from both experiments and
simulations is that the flow of the granular material remains in a
dynamic steady-state with the intruder for essentially the entire
trajectory, despite the highly transient nature of this process. By
dynamic steady-state, we mean that as the intruder moves through the
granular material, the flow field near the intruder scales linearly
with the instantaneous intruder speed, even as the intruder
decelerates rapidly. Since the force propagation speeds $v_f\sim 300 \gg v_0$~m/s~\cite{Clark2015} are
much faster than the intruder motion, forces can
propagate and relax fast enough that the motion of grains near the
intruder is essentially incompressible and remains in this dynamic
steady state. We expect our results to be applicable to a wide array
of rapid, highly transient dense granular flows, assuming $v_f\gg v_0$
(where $v_0$ sets a generic driving rate). Existing descriptions of
well developed granular flows~\cite{daCruz2005,Jop2006,Kamrin2012} may
be extended or modified~\cite{Dunatunga2015} to capture these
transient flows. Additionally, while force propagation depends crucially on the intergrain force law~\cite{Clark2015}, the agreement between flow field measurements in simulations and experiments is largely independent of the grain properties used in the simulations, suggesting that accurate grain-grain interactions are not necessary to model highly dynamic
flows, provided that $v_f\gg v_0$.

\begin{figure}
\raggedright (a) \\ \centering \includegraphics[trim=10mm 60mm 10mm 10mm, clip, width=0.8\columnwidth]{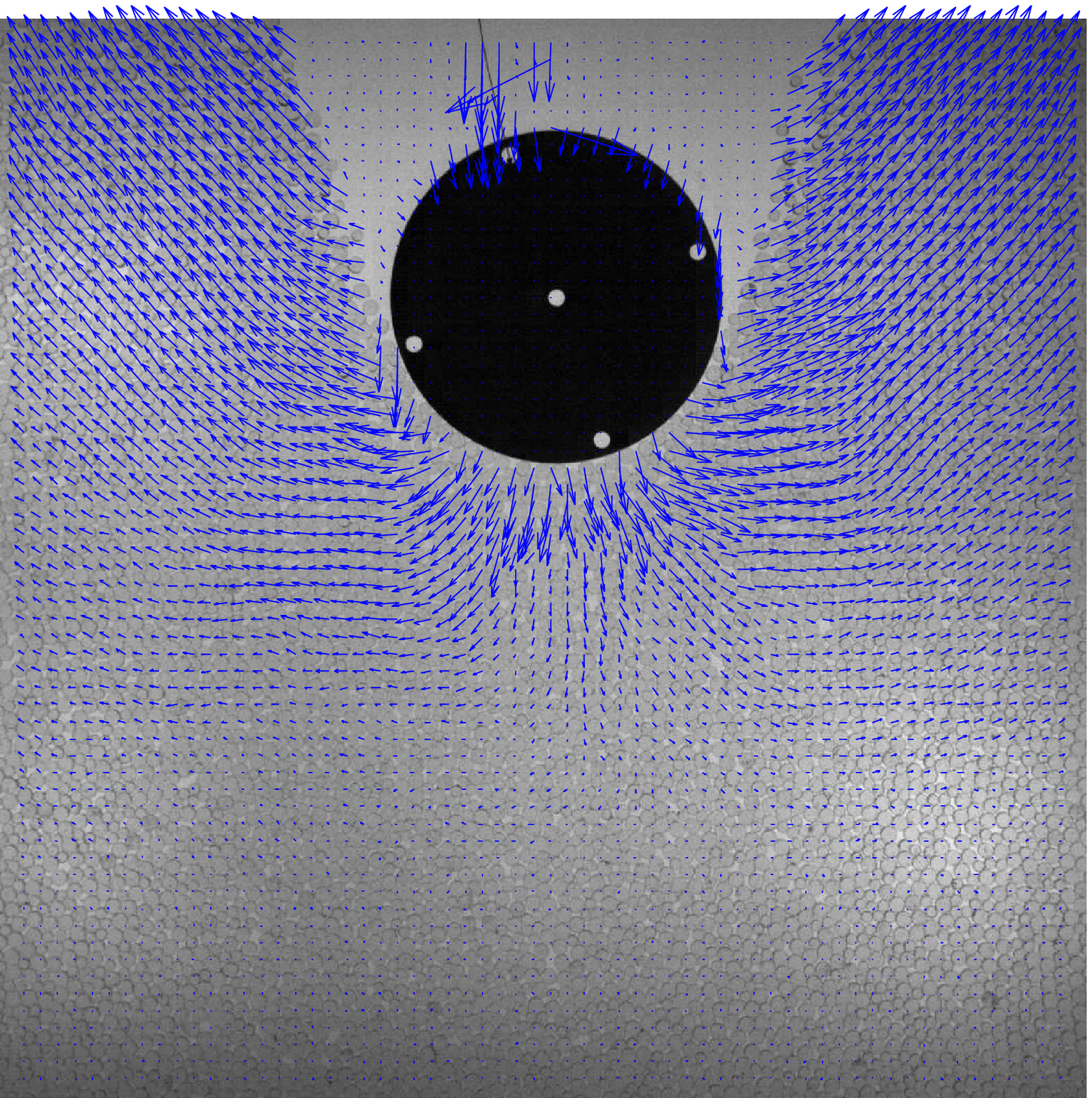}
\\ \raggedright (b) \\ \centering \includegraphics[trim=15mm 5mm 22mm 0mm,clip,width=0.9\columnwidth]{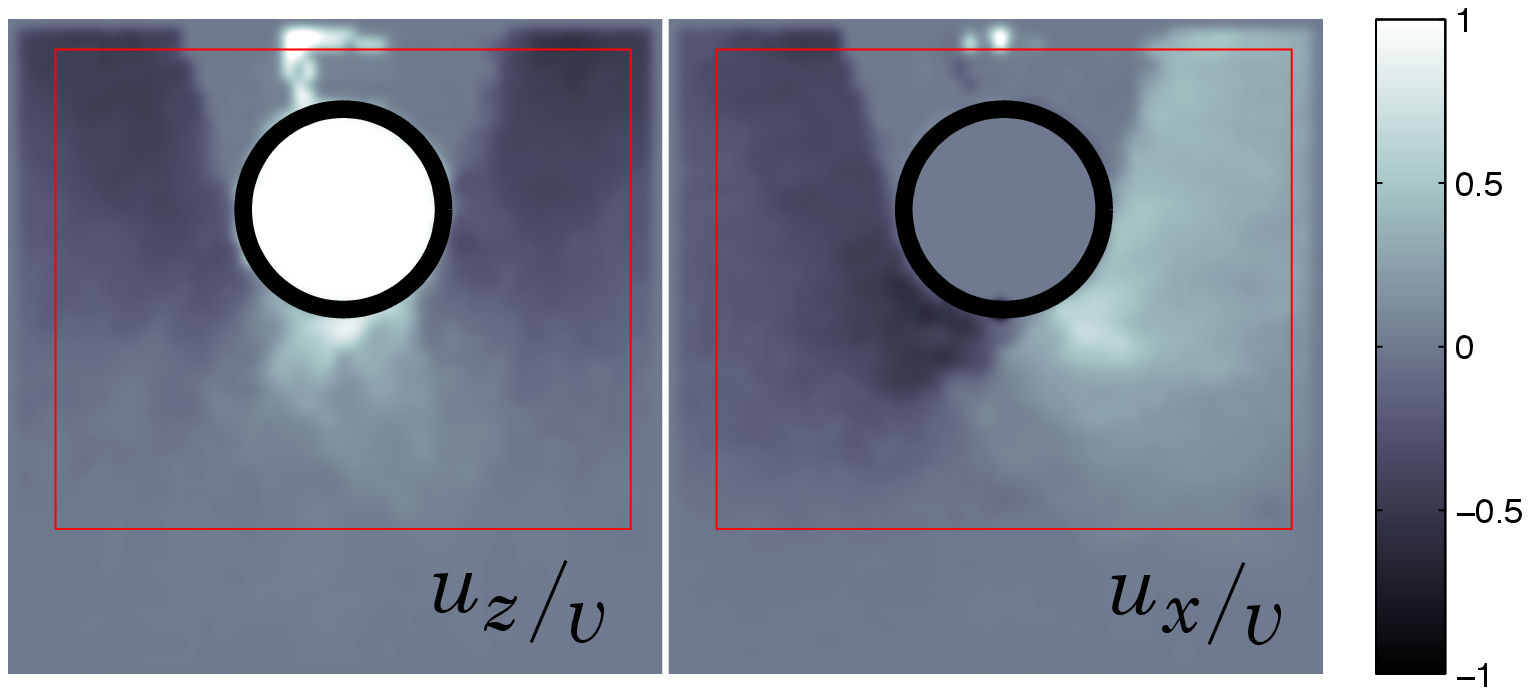}
\\ \raggedright (c) \\ \centering \includegraphics[trim=15mm 5mm 14mm 0mm,clip,width=0.9\columnwidth]{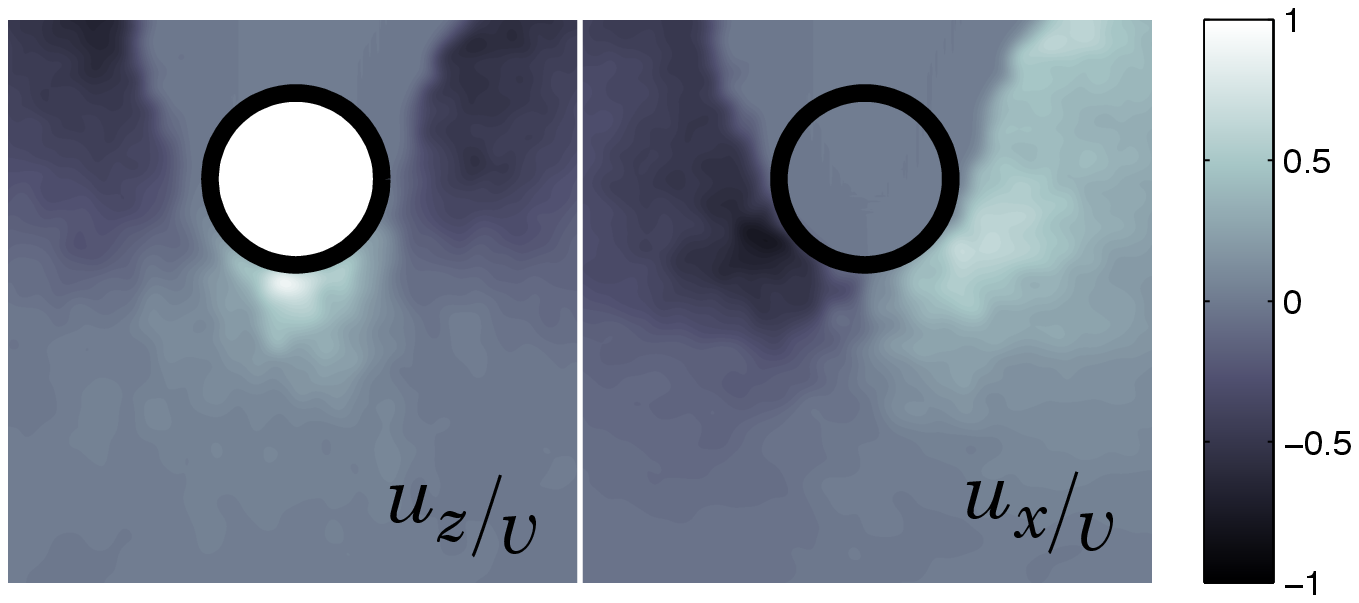}
\caption{(color online.) (a) PIV flow field for a circular intruder
  with radius $R=6.35$~cm at a particular frame. (b) Vertical and
  horizontal components of the spatially smoothed PIV flow field from
  experiment, normalized by the instantaneous intruder velocity
  ($v=3.05$~m/s). The left panel shows the normalized vertical
  velocity, $u_z/v$, with downward as positive, and the right panel
  shows the horizontal velocity, $u_x/v$, with rightward as
  positive. The white color in the intruder in the left panel denotes
  its downward motion. The red box encloses the region used to obtain
  the steady-state velocity field, which will move along with the
  intruder. (c) The instantaneous flow fields, $u_z/v$ and $u_x/v$, at
  a particular time from simulations with Hertzian, frictional
  interactions between grains.}
\label{fig:PIV}
\end{figure}

The experiments are carried out using the protocol described
in~\cite{Clark2012,Clark2013,Clark2014,Clark2015}. Here,
bronze intruders that are disks or have circular leading edges are
normally incident from above on photoelastic disks. We measure the granular flow fields using particle image
velocimetry (PIV) \cite{Grant01011997}, which analyzes successive pairs of frames from high-speed movies (sampled at 2333~Hz) to estimate the local flow field. This returns estimates of the local displacement on a grid, as shown in Fig.~\ref{fig:PIV}(a). The photoelastic disks are cut from PSM-1, manufactured by Vishay Precision Group. Here, $v_f\gg v_0$~\cite{Clark2015}, and we note that the grain-scale force picture and the subsequent intruder dynamics change drastically when $v_f \sim v_0$~\cite{pre12_impact,Clark2015}. However, when $v_f\gg v_0$, the intruder deceleration is dominated by large fluctuations in space and time in the form of quasi-random collisions with networks of particles that occur beneath the intruder~\cite{Clark2012,Clark2014}. Thus, our primary focus in this study is on the region directly beneath the intruder. The material responds quickly to the advancing intruder, and the fast force dynamics average over longer times to yield the rate-independent and Bagnold-like~\cite{Bagnold1954} velocity-squared
drag forces that are common in both impact studies~\cite{Allen1957,Forrestal1992,Katsuragi2007,Goldman2008,Goldman2010,Clark2012,Clark2013,Clark2014}
(\textit{i.e.}, transient driving) and steady drag
experiments~\cite{Albert1999,Albert2001,Geng2005,Seguin2011,Seguin2013,Takehara2010,Takehara2014}
(\textit{i.e.}, well-developed flows). We note that our results help explain
similarities between these two processes.

To explore the influence of the interaction force between the granular
particles, we carried out simulations using both linear and nonlinear
(Hertzian) force models that included friction as well as simulations with a linear force model and no interparticle friction. Simulations, described briefly below, are
similar to those discussed in~\cite{pre12_impact} but with a nonlinear force interaction
model as well as parameters that are matched to the experiments. We consider a rectangular domain in two dimensions with gravity. The domain size, as well as particle numbers and sizes are as in the
experiments. The particle-particle, particle-intruder, and
particle-wall interactions are modeled using the soft-sphere approach
that includes friction and particle rotations. We then solve the following (nondimensional) equations of motion for each particle (including the intruder):
\begin{eqnarray}
m_i\frac{d^2\mathbf{r}_i}{dt^2} &=& \mathbf{F_l}^n_{i,j}+    \mathbf{F_l}^t_{i,j}  +  m_i\mathbf{g},\nonumber
\\
I_i\frac{d\boldsymbol{\omega}_i}{dt} &=&
-\frac{1}{2}d_i\mathbf{n}\times\mathbf{F_l}^t_{i,j}.
\label{eq:motion}
\end{eqnarray}
For the linear force model, the  normal force is given by 
${\bf F_l}^n_{i,j} =  \left[ k_n x - \gamma_n \bar m {\bf v}_{i,j}\right ] {\bf n},$
where $r_{i,j} = |{\bf r}_{i,j}|$, ${\bf r}_{i,j} = {\bf r}_i - {\bf
  r}_j$, and the normal direction is defined by ${\bf n} = {\bf
  r}_{i,j}/r_{i,j}$. The compression is defined by $x =
d_{\rm ave}-r_{i,j}$, where $d_{\rm ave} = {(d_i + d_j)/2}$, $d_{i}$ and
$d_{j}$ are the diameters of the particles $i$ and $j$; ${\bf
  v}_{i,j}^n$ is the relative normal velocity. 

The nondimensional force constant $k_n$ is related to the dimensional
one, $k$, by $k = k_n mg/d$, where $m$ is the average particle mass,
$d$ is the average particle diameter, and $g$ is Earth's gravity. All
quantities are expressed using $d$ as the length scale, the binary
collision time, $\tau_c = \pi \sqrt{d/ 2 g k_n}$, as the time scale,
and $m$ as the mass scale. Then, $\bar m$ is the reduced mass, and
$\gamma_n$ is the damping coefficient related to the coefficient of
restitution, $e_n$, by $\gamma_n = -2\ln e_n/\tau_c$~, see,
\textit{e.g.},~\cite{kondic_99}. We take $e_n=0.5$ constant and ignore its
possible velocity dependence~\cite{schafer96}. The Hertzian
interaction model is implemented as ${\bf F}_h =\sqrt{d_i
  d_j/(d_i+d_j)} \sqrt{x} {\bf F}_l$. In principle, the force
constant could now be connected to the material properties of the
particles using the method described, \textit{e.g.}, in \cite{kondic_99}.
Instead, here we use the results of static tests carried out to
measure directly the functional relation between the normal force and
compression, see~\cite{Clark2015}. The normal force constant is then
found using the measured value of the force for $1\%$ compression.
The tangential force is computed using a standard Cundall-Strack
model~\cite{cundall79}; see \textit{e.g.}~\cite{pre12_impact} for the details
of implementation. The particle-particle and particle-intruder coefficient of friction is set to experimentally estimated value of $\mu = 0.8$~\cite{Puckett2013}; the particles making up the walls are made very
inelastic and frictional, with $\mu=0.9$ and $e_n=0.1$. The system is prepared by placing granular
particles on a rectangular lattice, with random size distribution of
the particles. The particles are given random initial velocities and
left to settle under gravity. Then, the whole domain is vibrated
gently to let the particles settle once more. The intensity of
vibrations does not appear to be important; we use $\Gamma =
a\omega^2/g$ ($a$ is the amplitude and $\omega$ frequency of
vibrations) in the range $[1,5]$ without any systematic change in the
results. We then place a circular intruder just above the bed with an initial downward velocity $v_0$.

Results from PIV (for experiments) or actual particle positions and velocities (for simulations) can be
spatially coarse-grained~\cite{Goldhirsch2002,Goldhirsch2010,Clark2012-0}, as shown in
Fig.~\ref{fig:PIV}(b) and (c), to give a continuum flow field
$\mathbf{u}(\mathbf{x},t)$, where $\mathbf{x}$ represents spatial
coordinates in the lab frame. The vertical component,
$u_z(\mathbf{x},t)$ (with downward being positive $z$), and the
horizontal component, $u_x(\mathbf{x},t)$ (with rightward being
positive $x$) components of the flow field at intruder speed
$v=3.05$~m/s are shown in Fig.~\ref{fig:PIV}(b) for experiments. The
grid size used for the PIV algorithm is approximately the same size as
a single particle, so the particle-scale fluctuations in the velocity
fields still persist. To compare simulation results to PIV, we use a
coarse-grained momentum field normalized by the average mass
density. (We normalize by the average mass density instead of a local
mass density field, since the coarse-grained, spatially varying mass
density field goes to zero at the free surface and near the intruder.)
This yields a flow field $\mathbf{u}(\mathbf{r},t)/v$, as shown in Fig.~\ref{fig:PIV}(c) from simulations for intruder radius $R=6.35$~cm.

\begin{figure}

	 \raggedright (a) \\ \centering \includegraphics[trim=15mm 5mm
      15mm 0mm,clip,width=\columnwidth]{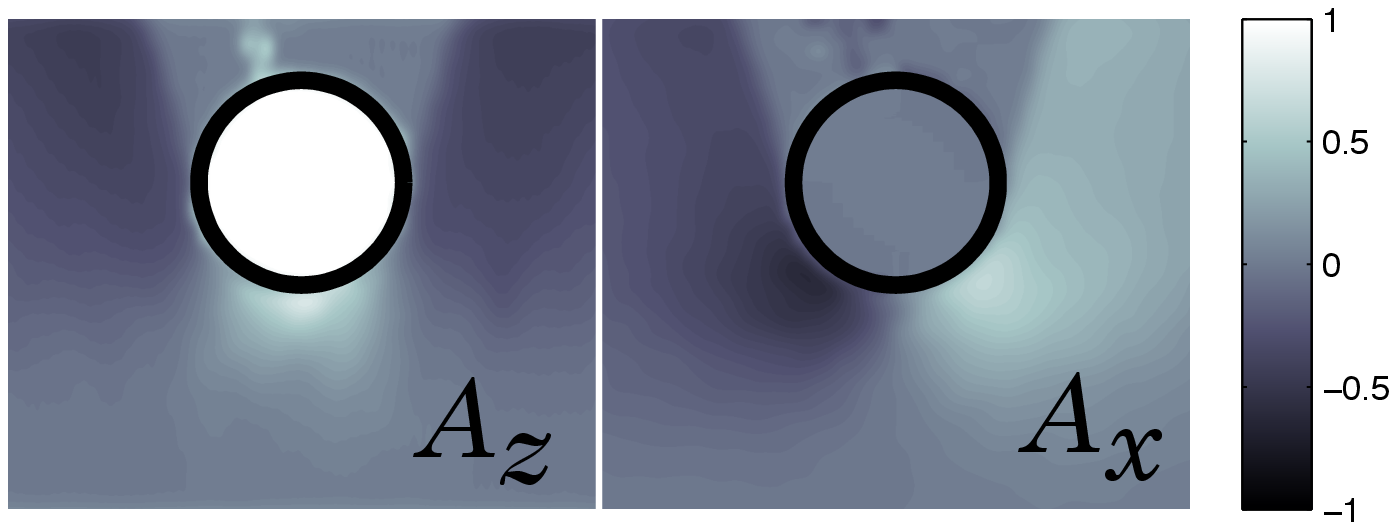}
    \\ \raggedright (b) \\ \centering \includegraphics[trim=15mm 5mm
      15mm 0mm,clip,width=\columnwidth]{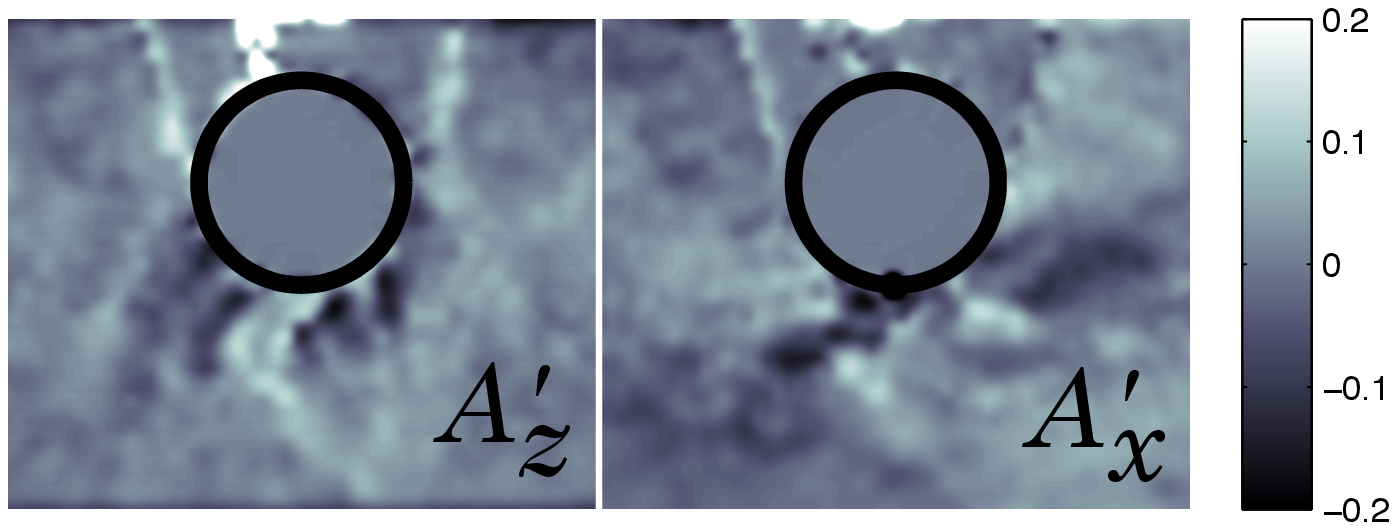}
    \\ \raggedright (c) \hspace{40mm} (d)
    \\  \includegraphics[trim=10mm 0mm 10mm 0mm, clip,width=.49\columnwidth]{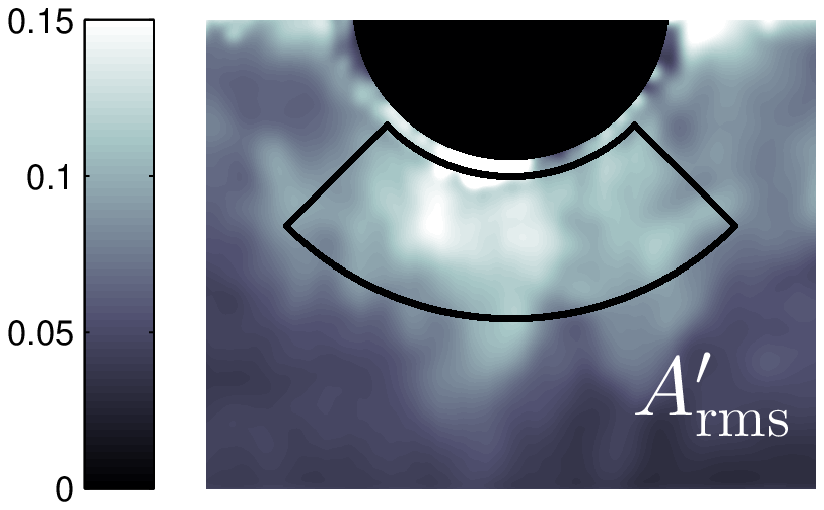} 
   		\includegraphics[trim=0mm 0mm 5mm 0mm, clip,width=.49\columnwidth]{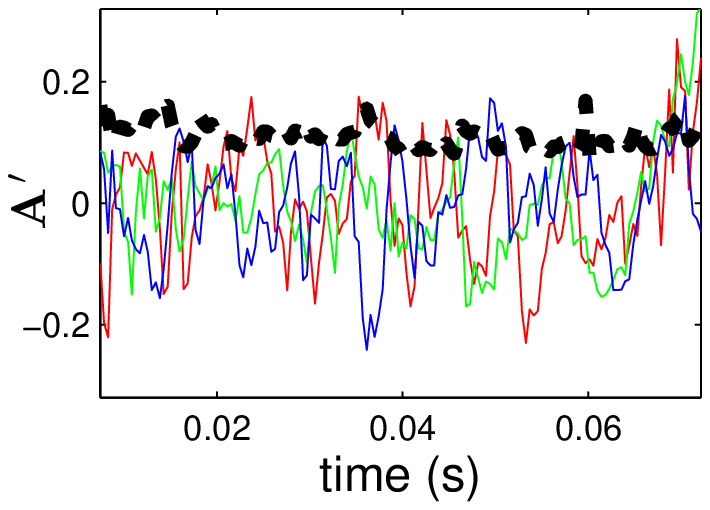}
      \caption{(color online.) (a) The average flow field, $\mathbf{A}$, from experiments for a
      circular intruder with radius $R=6.35$~cm. The left side shows
      the vertical velocity $A_z$ (down is positive), and the right
      side shows the horizontal velocity, $A_x$ (right is
      positive). (b) The instantaneous fluctuations, $A'_z$ and $A'_x$, in
      the flow field at a particular frame from experiments for a circular intruder
      with radius $R=6.35$~cm. (c) A spatial plot of $A'_{\rm rms}$, the root-mean-squared value of $\mathbf{A}'$. A time series of the spatial mean of $A'_{\rm rms}$ withiin the outlined region (where the fluctuations are most prominent) is plotted (thick dashed line) in panel (d). This quantity is essentially constant in time. Red, blue, and green solid lines in (d) show $A'_z$ at three points
      beneath the intruder. These signals fluctuate around zero with a
      correlation time of roughly 3~ms.}
\label{fig:avgCGvel}
\end{figure}

In both experiments and simulations, we find that 
\begin{equation}
\mathbf{u}(\mathbf{x},t)=v(t)\left[\mathbf{A}(\mathbf{x}-\mathbf{x}_0)+\mathbf{A}'(\mathbf{x}-\mathbf{x}_0,t)\right],
\label{eqn:steady-state-final}
\end{equation}
where $v(t)$ is the intruder speed (with motion assumed to be strictly downward), $\mathbf{x}_0(t)$ is the intruder position in the lab frame, $\mathbf{A}$ is the scaled steady-state velocity field, and $\mathbf{A}'$ captures the instantaneous fluctuations in the velocity field. $\mathbf{A}$ and $\mathbf{A}'$ are shown in Fig.~\ref{fig:avgCGvel}(a) and (b), respectively, for experiments. Similar fields for simulations with grain properties matched to those from the experiments (not shown) are indistinguishable by eye, and we quantitatively show that the two approaches agree in our analysis below. In each trajectory (experiments and simulations), we calculate $\mathbf{A}$ by averaging over many times using flow-field data inside the red rectangular region marked in Fig.~\ref{fig:PIV}(b). $\mathbf{A}$ appears very similar to the instantaneous flow fields shown in Fig.~\ref{fig:PIV}(b) and (c), but smoother spatially. $\mathbf{A}'$ is determined at each time from the difference between the instantaneous coarse-grained flow field and the normalized, space- and time-averaged flow field, $\mathbf{A}'=\mathbf{u}/v(t)-\mathbf{A}$. An experimental measurement of $\mathbf{A}'$ at one instant is shown in Fig.~\ref{fig:avgCGvel}~(b), which is typical for all times in both simulation and experiment.

\begin{figure}
\centering
 \includegraphics[trim=15mm 0mm 10mm 0mm,clip,width=\columnwidth]{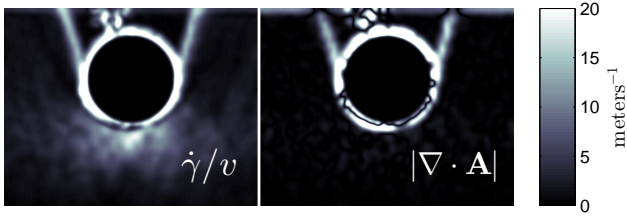}
\caption{The local shear rate $\dot{\gamma}/v$ and the divergence of $\mathbf{A}$, computed numerically as described in the text, showing that $\mathbf{A}$ is a divergence-free shear flow.}
\label{fig:div-and-shear}
\end{figure}

We find the fluctuations $\mathbf{A}'$ to be strongest beneath the
intruder, statistically stationary in time, and decoupled from the
intruder dynamics. Figure~\ref{fig:div-and-shear}(c) shows a spatial
plot of the root mean square (RMS) magnitude of the fluctuations,
$A'_{\rm rms}$. In the region beneath the intruder, the average fluctuations are
about $|\mathbf{A}'|\approx 0.1$. The magnitude $|\mathbf{A}'|$ is
always less than 0.2 (where a value of 1 would correspond to a local
velocity fluctuation of the same size as the intruder speed); it is
largest near the leading edge of the intruder and falls off rapidly
with increasing distance from the
intruder. Figure~\ref{fig:avgCGvel}(c) shows time-series plots of
$\mathbf{A}'$, which are statistically stationary in time.

By analyzing local strain rates, we find that $\mathbf{A}$ represents
a shear flow with zero divergence. Using numerical derivatives, we
compute the strain-rate tensor for the average flow field,
$\mathbf{D}=0.5[\nabla \mathbf{A}+(\nabla \mathbf{A})^\top]$, with
eigenvalues $d_1$ and $d_2$. Figure~\ref{fig:div-and-shear} shows the
local shear rate $\dot{\gamma}/v=(d_1-d_2)/2$ and demonstrates that $\nabla\cdot\mathbf{A}=\text{tr}\mathbf{D} =
  0$ within noise, \textit{i.e.} the flow of grains near the intruder is
  essentially incompressible. Note that $\dot{\gamma}/v$ is well
correlated to $A'_{\rm rms}(\mathbf{r})$, shown in
Fig.~\ref{fig:avgCGvel}(c). This is similar to many previous
studies~\cite{Menon1997,Losert2000}), where shear causes local
velocity fluctuations. Physically, $\mathbf{A}'$ represents non-affine
particle rearrangements as particles are forced to move past each
other, as opposed to a monotonic increase or decrease as the intruder
slows. A full analysis of grain-scale fluctuations, which could be
achieved with data for particle trajectories (as opposed to PIV), will
be a topic of future work.

Combined with force data from previous
studies~\cite{Clark2012,Clark2013,Clark2014}, the strain rates shown
in Fig.~\ref{fig:div-and-shear} can be used to estimate the inertial
number $I=\dot{\gamma}\sqrt{m/P}$, which is often used to determine a
constitutive relation for granular shear
flows~\cite{daCruz2005,Jop2006,Kamrin2012}. Here, $m$ is mass of a
single grain and $P$ is the local pressure. The maximum shear rate in
Fig.~\ref{fig:div-and-shear} is $\dot{\gamma}\approx 20v$ and the mass
of a grain $m$ is roughly 0.1~g. We estimate the pressure $P\sim F/D$
by considering the force $F$ on the intruder and dividing by the
intruder diameter $D$. $F$ is dominated by velocity-squared forces
which arise from collisions with force-chain-like
structures~\cite{Clark2014}, and, for circular intruders in the
present experiments, we find $F\approx h_0 v^2$, where $h_0$ is a shape and size dependent constant with units of kg/m; for the circular-nosed intruders considered here, we find $h_0/D\approx 5$~kg/m$^2$~\cite{Clark2014}. This yields $I\approx
0.09$ in the region directly beneath the intruder, which is in the
rapid flow regime, where nonlocal effects may be less important~\cite{Kamrin2012,Henann2013}.

\begin{figure}
\raggedright (a) \\ \centering \includegraphics[trim=0mm 0mm 0mm 0mm, clip, width=\columnwidth]{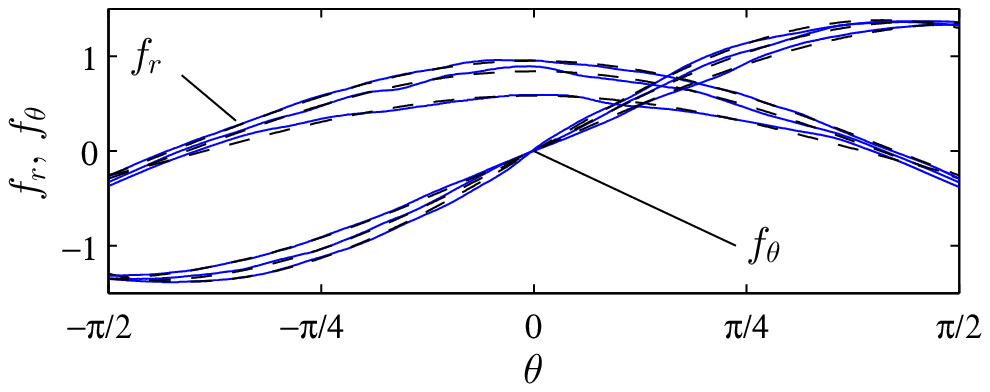}
\raggedright (b) \hspace*{37mm} (c)
\\ \centering \includegraphics[trim=0mm 0mm 0mm 0mm, width=0.49\columnwidth]{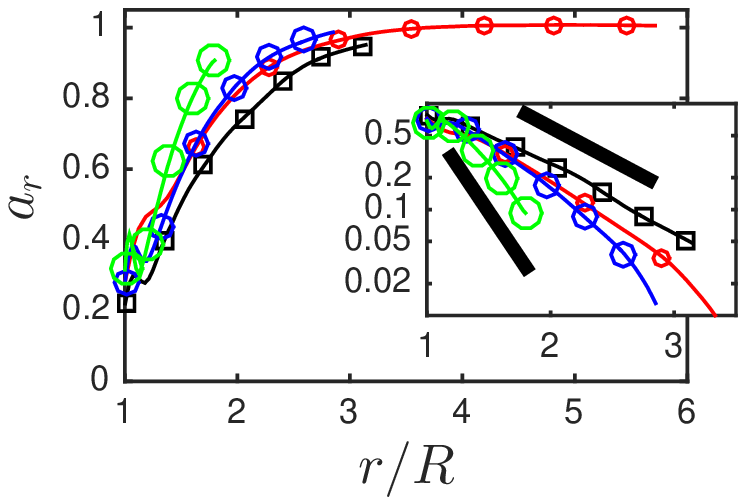}
\includegraphics[trim=0mm 0mm 0mm 0mm, width=0.49\columnwidth]{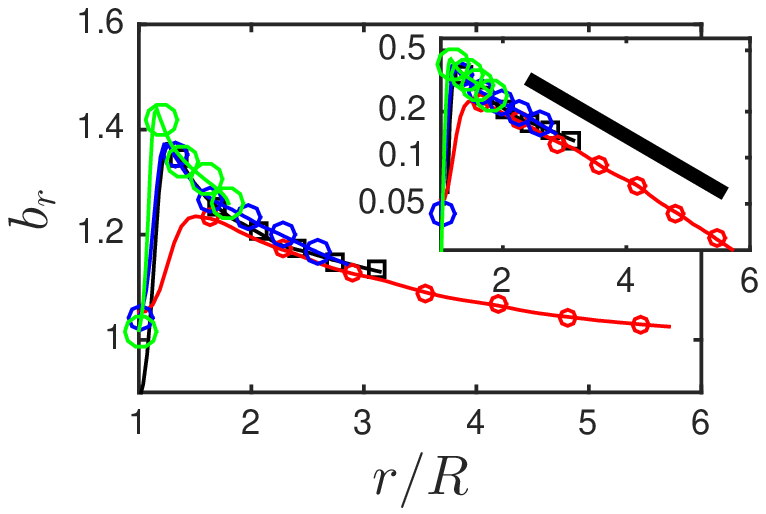}
\\ \raggedright (d) \hspace*{37mm} (e)
\\ \centering \includegraphics[trim=0mm 0mm 0mm 0mm, width=0.49\columnwidth]{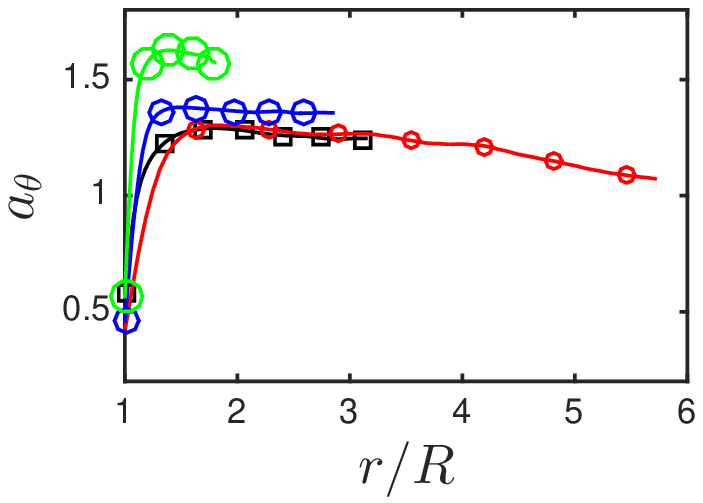}
\includegraphics[trim=0mm 0mm 0mm 0mm, width=0.49\columnwidth]{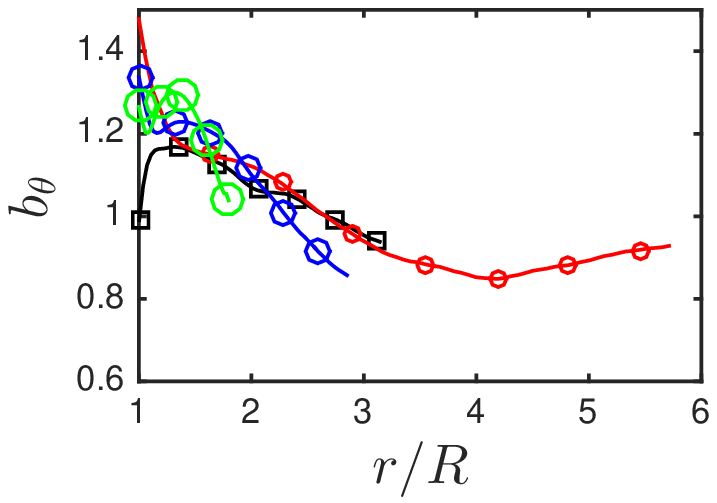}
\caption{(color online.) (a) Data (solid blue lines) and corresponding fit line (dashed black lines)
  of the form shown in Eq.~\eqref{eqn:flow-field-decomposition} for one intruder (radius $R=3.18$~cm) at $r=1.5R$,
  $2R$, and $3R$, where $r=R$ corresponds to the intruder
  boundary. (b)-(e) A comparison of the
  fit parameters---$a_r(r),b_r(r),a_\theta(r),b_\theta(r)$---for
  circular intruders with $R=3.18$~cm (small red circles) $R=6.35$~cm (medium blue
  circles), $R=10.15$~cm (large green circles), as well as the circular
  nosed intruder with $R=4.65$~cm and a rectangular tail (black
  squares). The inset of (b) shows semi-log plots of $1-a_r(r)$ versus
  $r/R$; thick black reference lines show exponential
  decay with decay lengths of $0.7R$ (upper) and $0.25R$ (lower). The
  inset of (c) shows semi-log plots of $b_r(r)-1$ versus $r/R$;
  thick black reference line shows exponential decay with
  decay length $1.85R$.}
\label{fig:flow-field-fits}
\end{figure}

To quantitatively compare $\mathbf{A}$ for various intruder sizes and simulation settings, we fit $\mathbf{A}$ to a functional form by decomposing it into radial and angular components
\begin{equation}
\mathbf{A}(\mathbf{x})=\mathbf{\hat{r}} \left[\cos\theta - f_r(\mathbf{r})\right] + \boldsymbol{\hat{\theta}} \left[f_\theta(\mathbf{r})-\sin\theta\right].
\label{eqn:flow-field-decomposition}
\end{equation}
Here, $\mathbf{r}=r\mathbf{\hat{r}}+ \theta\boldsymbol{\hat{\theta}}$,
where $\mathbf{r}=0$ corresponds to the center of the intruder,
$\theta$ is measured counterclockwise from the (downward) $z$-axis. The components $f_r$ and $f_\theta$ represent the flow field components in the intruder frame, and shifting by
$\mathbf{\hat{z}} = \cos\theta \mathbf{\hat{r}}-\sin\theta
\boldsymbol{\hat{\theta}}$ transfers these components back to the lab
frame, where $\mathbf{A}$ is defined. $f_r$ and $f_\theta$ are defined as
\begin{align}
f_r(\mathbf{r})= a_r(r)\cos[b_r(r)\theta]\\
f_\theta(\mathbf{r})= a_\theta(r)\sin[b_\theta(r)\theta].
\label{eqn:flow-field-decomposition-2}
\end{align}
Seguin et al. \cite{Seguin2011,Seguin2013} used a similar form to
describe quasistatic granular flow around downward-moving circular
obstacles, but with $b_r=b_\theta =1$, since, in their quasi-static
case, the flow was symmetric ahead of and behind the intruder. Here,
we consider fits only to the half-space in front of the
intruder. Sample fits at particular values of $r$ are shown in
Fig.~\ref{fig:flow-field-fits}. Far away, all four fit parameters
should approach $1$, corresponding to no grain motion.

Figure~\ref{fig:flow-field-fits} shows $a_r(r)$, $b_r(r)$,
$a_\theta(r)$, and $b_\theta(r)$ for different circular-nosed
intruders with radii $R=3.18$, 4.65, 6.35, and 10.15~cm. The intruder
with $R=4.65$~cm is an ogive, with a circular nose and rectangular
tail; however, the particles are never in contact with the tail, so
that its presence is irrelevant, aside from increasing the area of the
intruder and therefore its mass. The fit parameters for each intruder appear
similar when rescaled by $R$, with secondary dependencies on the ratio
$\rho_{\rm int}/\rho_g$ of intruder to grain mass density and on the
ratio $d/R$ of grain size to intruder radius, where $d\approx 5$~mm.
$a_r$ and $b_r$ decay roughly exponentially to their far-field values
as $a_r \propto \exp(-r/\xi_{a_r})$ and $b_r \propto
\exp(-r/\xi_{b_r})$, with $0.25R < \xi_{a_r}< 0.7R$ and $\xi_{b_r}
\approx 1.85 R$. This localization and exponential spatial decay is
also common in dense granular flows that are driven by a boundary
(\textit{e.g.}, Refs.~\cite{Losert2000,Seguin2013,Seguin2011}, and many
others).

\begin{figure}
\raggedright (a) \hspace*{37mm} (b)
\\ \centering \includegraphics[trim=0mm 0mm 0mm 0mm, width=0.49\columnwidth]{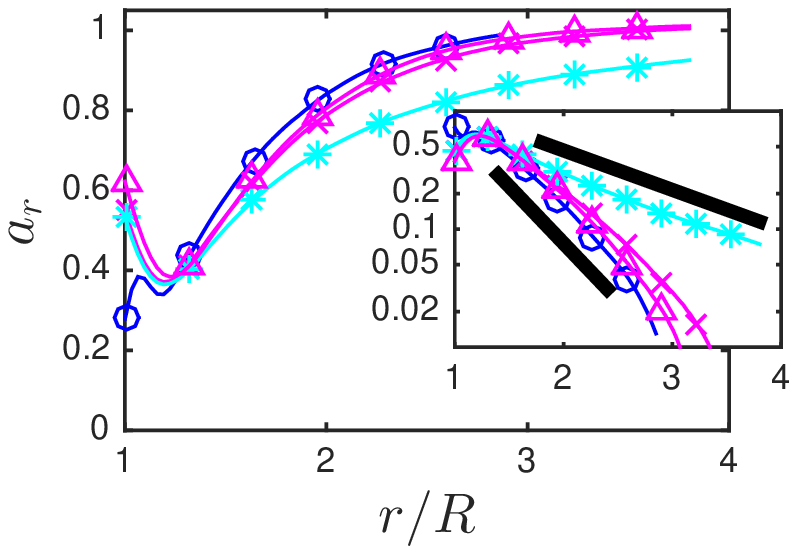}
\includegraphics[trim=0mm 0mm 0mm 0mm, width=0.49\columnwidth]{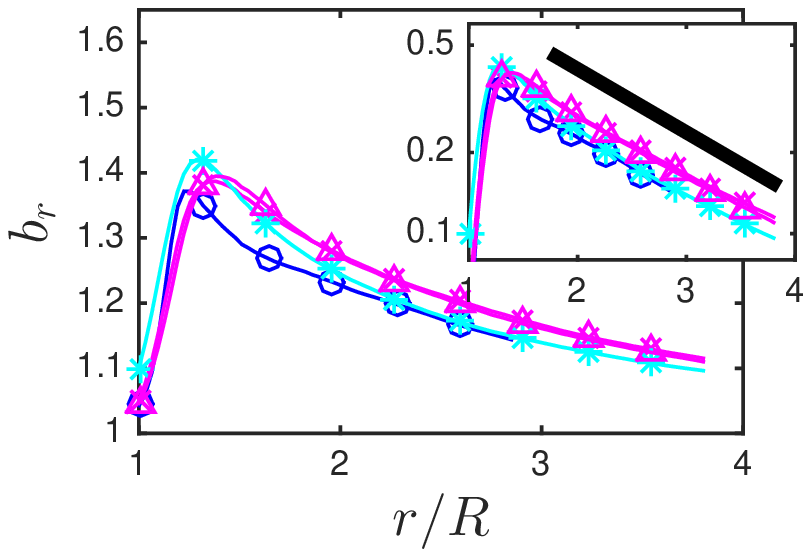}
\\ \raggedright (c) \hspace*{37mm} (d)
\\ \centering \includegraphics[trim=0mm 0mm 0mm 0mm, width=0.49\columnwidth]{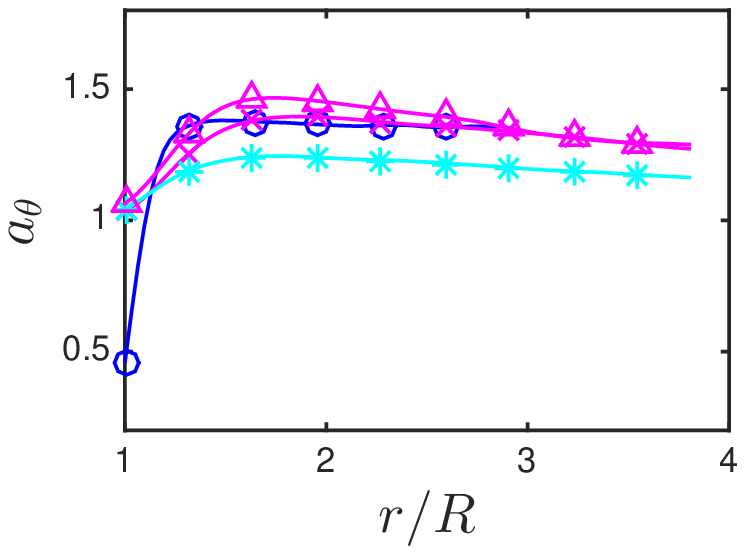}
\includegraphics[trim=0mm 0mm 0mm 0mm, width=0.49\columnwidth]{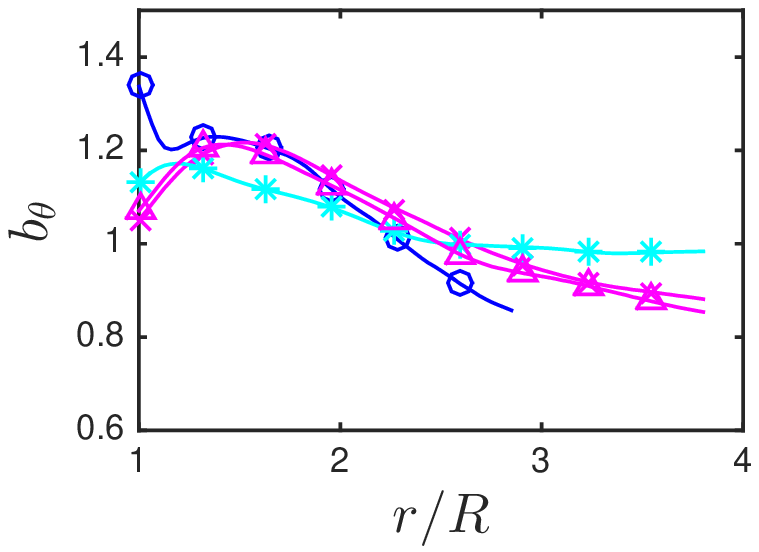}
\caption{(color online.) Comparison of the parameters $a_r(r)$, $b_r(r)$, $a_r(r)$, and $a_\theta(r)$ between experimental data (blue circles) from Fig.~\ref{fig:flow-field-fits} and simulations using frictional linear (magenta crosses), frictional Hertzian (magenta triangles) and frictionless linear (light blue asterisks) interactions, all under the same conditions with $R=6.35$~cm. Insets of (a) and (b) show semi-log plots of $1-a_r(r)$ and $b_r(r)-1$, respectively, versus $r/R$. Thick black reference lines in the insets show exponential decay with decay lengths of (a) of $1.3R$ (upper) and $0.45R$ (lower), and in (b) of $1.85R$.}
\label{fig:flow-field-fits-compare}
\end{figure}

Figure~\ref{fig:flow-field-fits-compare} shows a comparison between the fit of experimental and computational data to Eqs.~\eqref{eqn:flow-field-decomposition}-\eqref{eqn:flow-field-decomposition-2} for a single size intruder; this is typical for all sizes. We find surprisingly good agreement for all considered interparticle force models (frictional Hertzian, frictional linear, and frictionless linear interactions). Such a good agreement shows that in the present regime where $v_0 \ll v_f$, the details of the force model are not crucial for the response of granular material. However, we note that frictional forces primarily affect the decay length of the radial flow field, as shown in the inset of Fig.~\ref{fig:flow-field-fits-compare}(a), and thus they play an important role in determining dynamics of the intruder; in particular, without friction, the final penetration depth is almost 50\% larger (roughly 60~cm) than when friction is present (roughly 42~cm). The functional form of the normal forces between the particles (linear versus nonlinear) however does not appear to be important.

These results provide several important physical insights that should
be applicable to a broad class of shear-like flows that are both rapid
and highly transient, but where driving speed, which is here set by
$v(t) \le v_0$, is still very slow compared to the granular force
transmission speed $v_f$. In the granular flow fields, we observe none
of the elastic-like response (\textit{i.e.}, loading and unloading) that is
dominant when $v_0\sim v_f$~\cite{pre12_impact,Clark2015}. Instead, we
observe that the particle motion scales linearly with driving speed,
which also occurs for well-developed shear flows in the limits of both
small (quasi-static) and large (rapid driving) inertial number
$I$, with a transition region in
between~\cite{Koval2009,Kamrin2012}. Our system is clearly more akin
to the limit of large $I$, with $I \sim 10^{-1}$ at the leading edge
of the intruder, but descriptions of such flows explicitly exclude
transients in the driving speed. However, we observe a dynamic steady
state of the granular flow during highly transient driving, which
suggests that rapid, highly transient granular flows may fall into the
same class as well-developed rapid flows, provided $v_f\gg
v_0$. Conversely, it is possible that models such as $\mu(I)$ could be
extended to processes such as granular impact where the flow is
transient. For example, a recent study~\cite{Dunatunga2015} presents a
modified $\mu(I)$ rheology to study dynamic granular flows, and our
results here suggest that this approach will likely be successful in
many cases. In addition, simulations show that, although final
penetration depth is strongly influenced by frictional interactions,
the granular flow in this regime appears relatively insensitive to the
form of the grain-grain force law (\textit{e.g.}, linear versus Hertzian,
consistent with~\cite{Campbell2002}) or even to the presence of
friction~\cite{Seguin2009,Clark2014}.

\begin{acknowledgments}
This work has been supported by the U.S. DTRA under Grant
No. HDTRA1-10-0021, by NASA grant NNX15AD38G, and by NSF Grant
DMR-1206351.
\end{acknowledgments}

\bibliography{References}

\end{document}